\documentstyle[12pt]{article}
\begin{document}
\baselineskip=16pt

\title{Diffusion process in a flow}
 
\author{Piotr Garbaczewski\\
Institute of Theoretical Physics, University of Wroc{\l}aw,
pl. M. Borna 9, \\
PL-50 204 Wroc{\l}aw, Poland}

\maketitle
\hspace*{1cm}

\begin{abstract}
We establish  circumstances  under which  the  dispersion of passive
contaminants in a forced, deterministic or random, flow can be
consistently
interpreted as a Markovian diffusion process.
In case of conservative forcing the
repulsive  case only, $\vec{F}=\vec{\nabla }V$ with $V(\vec{x},t)$
bounded from below,  is  unquestionably admitted by the
compatibility conditions.
A class of diffusion processes is exemplified, such that the
attractive
forcing is  allowed as well,  due to an appropriate compensation
coming from the "pressure"   term. The compressible Euler flows
form their subclass, when regarded as stochastic processes.
\end{abstract}
\vskip1.0cm

Whenever one tries to  analyze  random
perturbations that are either superimposed upon or intrinsic to a 
driving deterministic  motion, quite typically
a configuration space equation
$\dot{\vec{x}}=\vec{v}(\vec{x},t)$ is invoked,  which  is next
replaced by
a formal infinitesimal representation  of   an It\^{o} diffusion 
process
 $d\vec{X}(t)= \vec{b}(\vec{X}(t),t)dt + \sqrt{2D} d\vec{W}(t)$. 
Here,  $\vec{W}(t)$  stands for  the normalised  Wiener noise, and  
$D$  for a diffusion constant.  

The dynamical meaning of $\vec{b}(\vec{x},t)$
 relies on a  specific  diffusion input and  its possible phase-space
 (e.g. Langevin)  implementation,  that  entail  a detailed
 functional
relationship of $\vec{v}(\vec{x},t)$ and $\vec{b}(\vec{x},t)$,
and justify
  such notions like:  diffusion in an external force field,
  diffusion along,
     against or across the deterministic flow, \cite{horst},
     also with shear effects, \cite{horst1}.

The pertinent mathematical formalism  corroborates 
both the Brownian motion of a single particle and the diffusive
transport 
of neutrally buoyant components in flows of the hydrodynamic type.

Clearly, in random media that are 
statistically at rest, diffusion of single tracers or dispersion of 
pollutants are well described by the Fickian outcome of the molecular 
agitation, also in the presence of external force fields (then  
in terms of Smoluchowski diffusions).  On the other hand, it is
of fundamental importance to understand how statistically relevant
flows 
in a random medium  (fluid, as example) affect dispersion.
In the context
of  fluids, we might   refer to diffusion enhancement due to
turbulence,
behaviour of Brownian particles in shear flows, but also to general
effects of  the external forcing (various forms of deterministic  or  
random "stirring" of the random medium) exerted  upon gradient
or  non-gradient, compressible and  incompressible
flows,  and carried by them passive constituents, \cite{horst1}.

Except for suitable continuity and growth restrictions, necessary to
guarrantee  the existence  of the process $\vec{X}(t)$ governed by
 the It\^{o} stochastic
differential equation, the choice of the driving velocity field
$\vec{v}(\vec{x},t)$ and hence of the related drift
$\vec{b}(\vec{x},t)$ is normally regarded   to be arbitrary.

However, the  situation looks otherwise, \cite{horst1},  if  we are
interested in a diffusion of passive tracers in the  a priori given
flow  whose velocity field is  a solution of the nonlinear
partial differential equation, be it Euler,
Navier-Stokes, Burgers or the like.
An implicit  assumption, that passively buoyant in a fluid tracers
have a
negligible effect  on the flow, looks  acceptable (basically,
in case when
the concentration of a passive component in a flow is small).
Then,  one is tempted   to view  directly
 the fluid velocity field $\vec{v}(\vec{x},t)$
as  the forward drift $\vec{b}(\vec{x},t)$ of the process,
with the contaminant
being diffusively dispersed along the streamlines.

However,  in  general,  the assumed nonlinear evolution rule for
$\vec{v}(\vec{x},t)$ must be checked against the dynamics
that is allowed to govern the space-time dependence of the
forward drift
field $\vec{b}(\vec{x},t)$, \cite{nel}, which  is \it not \rm at all
arbitrary. 
The  latter is ruled by  standard consistency  conditions 
that are respected by any 
 Markovian diffusion process, and additionally by  the rules of the 
forward and backward It\^{o} calculus, \cite{horst,nel}.

This particular issue we have analyzed 
before in the context of Burgers flows, \cite{burg}, where
the Burgers velocity field was found to be  inappropriate to stand for
 the forward drift of a Markovian diffusion  process.
Actually, the backward  drift was
a correct identification. Then, the  forced Burgers dynamics
$${\partial _t\vec{v}_B + (\vec{v}_B\cdot \vec{\nabla })\vec{v}_B=
D \triangle \vec{v}_B +
\vec{\nabla }\Omega }\eqno (1)$$
and  the diffusion-convection equation
$${\partial _t c +
(\vec{v}_B\cdot \vec{\nabla })c = D\triangle c}\eqno (2)$$
 for  the concentration
$c(\vec{x},t)$ of a passive  component in a flow,
in case of gradient velocity fields, were proved to be
compatible with the Markovian diffusion process input.

According to Ref. \cite{burg}, in that case the dynamics of
concentration results from
 the  stochastic diffusion process whose density $\rho (\vec{x},t)$
evolves according to
$${\partial _t \rho  = - D\triangle \rho -
\vec{\nabla }
\cdot (\vec{v}_B\rho )\enspace ,}\eqno (3)$$
 or equivalently:
 $${\partial _t \rho =
 D\triangle  \rho - \vec{\nabla }\cdot (\vec{b}\rho )\enspace ,}
 \eqno (4)$$
 $$\vec{b} \doteq
 \vec{v}_B +  2D\vec{\nabla }ln \rho \enspace .$$

The previous reasoning can be easily exemplified
by considering   the standard unforced Brownian motion with the
initial
(arbitrary, but sufficiently regular) density $\rho _0(\vec{x})$.
Its evolution
$\rho _0(\vec{x}) \rightarrow \rho (\vec{x},t)$ is
implemented  by the conventional heat kernel $p(\vec{y},s,\vec{x},t)=
[4\pi D(t-s)]^{-1/2} exp[-{(x-y)^2\over {4D (t-s)}}]$.
The backward drift
of the process (a solution of the unforced Burgers equation,
 originally denoted
$\vec{b}_*(\vec{x},t)$ in Ref. \cite{burg}) is defined
as follows:  $\vec{v}_B(\vec{x},t)=-2D\vec{\nabla }ln \rho $.
The  pertinent
concentration dynamics is given by
$${c(\vec{x},t)=
\int p_*(\vec{y},0,\vec{x},t) c_0(\vec{y}) d^3y}\eqno (5)$$
$$p_*(\vec{y},0,\vec{x},t) \doteq  p(\vec{y},0,\vec{x},t)
{\rho _0(\vec{y})\over {\rho (\vec{x},t)}}\enspace .$$

The remaining part is to determine the
function $c_0(\vec{x},t)$ i.e. the concentration of a tagged
population in a Brownian ensemble.
If we arbitrarily decompose the density of the process
into $\rho =\rho _1 + \rho _2$ and regard $\rho _1(\vec{x},t)$ as the
density of a
tagged    population, then an appropriate  definition of the
concentration comes  through:
$${c(\vec{x},t)={{\rho _1(\vec{x},t)}\over {\rho (\vec{x},t)}}
\enspace .}
\eqno (6)$$

By inspection one can check the validity of the diffusion-convection
equation
for $c(\vec{x},t)$ in a  Brownian flow with the (backward drift)
velocity  $\vec{v}_B(\vec{x},t)=-2D\vec{\nabla }ln \rho $.

By combining  intuitions which underly the self-diffusion
description,
\cite{spohn},  with those appropriate for  probabilistic solutions
of the
so-called Schr\"{o}dinger boundary-data and next-interpolation
problem, \cite{burg,olk,zambr},
 the above argument can be generalized to
conservatively forced  diffusion processes.

Namely, let us consider again the density
$\rho (\vec{x},t), t\geq 0$
of a stochastic diffusion process, solving   the Fokker-Planck
equation
$\partial _t\rho = D\triangle  \rho -
\vec{\nabla }\cdot (\vec{b} \rho )$,
where $\vec{b}(\vec{x},t)$ stands for a forward drift.
In case of  conservative forcing, the drift solves an
evolution equation:
$${\partial _t\vec{b} +(\vec{b}\cdot \vec{\nabla })\vec{b} =
- D\triangle \vec{b}
+ \vec{\nabla } \Omega \enspace .}\eqno (7)$$

  For drifts that are gradient fields, the potential
$\Omega $, \it whatever \rm its functional form is,
\it must \rm allow for  a  representation  formula, reminiscent of
the probabilistic Cameron-Martin-Girsanov transformation:
$${\Omega (\vec{x},t) = 2D[ \partial _t\Phi + {1\over 2}
({\vec{b}^2\over {2D}} + \vec{\nabla }\cdot \vec{b})]\enspace ,}
\eqno (8)$$
 where $\vec{b}(\vec{x},t) = \vec{\nabla } \Phi (\vec{x},t)$.

For the existence of
the Markovian diffusion process with the forward drift
$\vec{b}(\vec{x},t)$, we must  resort to  potentials
$\Omega (\vec{x},t)$ that are  \it not  \rm completely  arbitrary
functions.
Technically, \cite{olk}, the minimal requirement is that the
potential is
 bounded from below.
 This restriction will have profound consequences for
 our further discussion of diffusion in a flow.

If we set  $\rho = \rho _1 + \rho _2$ again, and demand that
$\rho _1 \neq \rho $  solves the Fokker-Planck equation with
the very
same drift  $\vec{b}(\vec{x},t)$ as $\rho $ does, then as a
necessary consequence of the general
formalism, \cite{burg,olk},  the concentration
$c(\vec{x},t)={{\rho _1(\vec{x},t)}\over {\rho (\vec{x},t)}}$
solves an associated diffusion-convection equation $\partial _t c +
(\vec{v}_B\cdot \vec{\nabla })c= D\triangle c$.
Here,  the flow velocity
$\vec{v}_B(\vec{x},t)$ coincides with the backward drift
$\vec{b}_*\dot= \vec{v}_B$ of the generic  diffusion process
with the   density $\rho (\vec{x},t)$  and reads:  $\vec{v}_B =
\vec{b} - 2D \vec{\nabla } ln  \rho $.
Obviously, the forced Burgers
equation  (1) is identically satisfied.

We should clearly discriminate between forces whose effect is a
"stirring" of the random medium and those acting selectively on
diffusing
 particles, with a negligible effect on the medium itself.
 For example, the traditional Smoluchowski diffusion processes  in
  conservative force fields  are considered  in random media that are
  statistically at rest.  Following the standard (phase-space,
  Langevin) methodology, let  us
  set $\vec{b}(\vec{x})={1\over \beta }\vec{K}(\vec{x})$,
  where $\beta $
  is a (large) friction coefficient  and $\vec{K}$ represents an
   external Newtonian force per unit of mass ( e.g. an acceleration)
  that is of gradient from, $\vec{K}=-\vec{\nabla }U$.
  Then, the effective
  potential $\Omega $ reads:
  $${\Omega  = {\vec{K}^2\over {2\beta ^2}}+ {D\over \beta }
  \vec{\nabla }\cdot \vec{K}}\eqno (9)$$
and the only distinction  between the
  attractive or repulsive cases  can be read out from
     the term  $\vec{\nabla }\cdot \vec{K}$.
  For example, the harmonic attraction/repulsion
  $\vec{K}=\mp  \alpha \vec{x}, \, \alpha >0$ would give rise to
  a harmonic  repulsion, if interpreted in terms of
   $\vec{\nabla }\Omega$, in view of
  $\Omega = {\alpha ^2\over
  {2\beta ^2}}\vec{x}^2 \mp  3D{\alpha \over \beta}$. The innocent
  looking $\mp 3D{\alpha \over \beta }$ renormalisation of the
   quadratic function gives rise to entirely
  different diffusion processes, with an equilibrium measure arising
  in case of $U(\vec{x})= + {\alpha \over 2} \vec{x}^2$ only.

The situation would not  change under  the
incompressibility condition (cf. also the probabilistic approaches to
the Euler, Navier-Stokes and Boltzmann equations, \cite{marra}).
Following  Townsend' s, \cite{horst1},
 early investigation of the diffusion of heat spots in isotropic
 turbulence we may choose
 $U(\vec{x})= {\alpha \over 2}x^2 - {\alpha
 \over 4} (y^2 + z^2)$ which implies $\vec{\nabla }\cdot \vec{K} = 0$.
Then, $\Omega (\vec{x}) = {\alpha ^2\over {2\beta ^2}}[x^2 + {1\over
4}(y^2 + z^2)]$, hence the repulsive $\Omega $ is produced again in
the
equation of motion characterising a stationary diffusion in
an incompressible fluid: $div\, \vec{v}=0$,
$\vec{b}=\vec{b}_*=\vec{v} \rightarrow  (\vec{v}\cdot \nabla )
\vec{v} =
{\vec{\nabla } \Omega }$.

By formally changing a sign of $\Omega $ we would arrive at
the attractive variant of the problem, that is however incompatible
with the diffusion process scenario in view of the unboundedness of
$- \Omega $ from below.

We have thus arrived at the major point of our discussion:
we may  get in
trouble with the Markovian diffusion input in case of
 general  external "stirring" forces. Hence,  we must specify
 an admissible class of  perturbations which,  while
modifying the flow dynamics, would nonetheless generate a consistent
diffusion-in-a-flow  transport of passive tracers.

Should we a priori exclude the attractive variants of the potential
$\Omega $ ?  Can we save the situation by incorporating,
hitherto not considered, "pressure" term effects as suggested by
the general form of the compressible Euler (here
 $\vec{F}=-\vec{\nabla }V$ stands for external volume forces and
 $\rho $ for the fluid density that itself undergoes a stochastic
 diffusion process):
$${\partial _t\vec{v}_E + (\vec{v}_E\cdot \vec{\nabla })\vec{v}_E =
\vec{F} - {1\over {\rho }}\vec{\nabla }P}\eqno (10)$$
or  the incompressible, \cite{marra}, Navier-Stokes equation:
$${\partial _t\vec{v}_{NS} + (\vec{v}_{NS}\cdot \vec{\nabla })
\vec{v}_{NS} =  {\nu \over {\rho }}\triangle \vec{v}_{NS} + \vec{F} -
{1\over{\rho }}\vec{\nabla }P\enspace ,}\eqno (11)$$
both to be compared with the equations (1) and (7), that set
 dynamical constraints  for respectively backward and forward drifts
 of a Markovian diffusion process ?

Notice that the acceleration term $\vec{F}$ in equations (10) and (11)
normally is regarded   as arbitrary, while the corresponding term
$\vec{\nabla }\Omega $   in (1) and (7)  involves a bounded from
below function    $\Omega (\vec{x},t)$.

Since, in case of gradient velocity fields, the dissipation term
in the incompressible  Navier-Stokes equation (11) identically
vanishes,
we should concentrate on analyzing the possible
"forward drift of the Markovian process"  meaning of the Euler
flow with the velocity field   $\vec{v}_E$,  (10).

At this point it is useful, at least on the formal grouds, to invoke
the standard phase-space argument that is valid for a Markovian
diffusion process
taking place in a given flow  $\vec{v}(\vec{x},t)$ with as yet
unspecified dynamics. We account for an explicit force exerted upon
diffusing particles, while not necessarily directly
affecting the driving flow itself.
Namely, \cite{horst1,nel}, let us  set for infinitesimal increments of
phase space random variables:
$$d\vec{X}(t)= \vec{V}(t) dt $$
$${d\vec{V}(t)= \beta [\vec{v}(\vec{x},t) - \vec{V}(t)] dt +
\vec{K}(\vec{x})dt  + \beta \sqrt{2D} d\vec{W}(t)\enspace .}
\eqno (12)$$

Following the leading idea of the Smoluchowski approximation, we
assume
 that $\beta $ is large, and consider the process for  times
 significantly exceeding $\beta ^{-1}$. Then, an appropriate
  choice of
 the velocity field $\vec{v}(\vec{x},t)$
 (boundedness and  growth restrictions  are  involved) may
 in principle guarrantee, \cite{nel},  the convergence of the
 spatial part
 $\vec{X}(t)$ of the process (12) to  the It\^{o} diffusion process
 with infinitesimal increments:
 $${d\vec{X}(t) = \vec{v}(\vec{x},t)dt + \sqrt{2D}
 d\vec{W}(t)\enspace .}\eqno (13)$$

However, one cannot blindly insert in the place of the forward drift
$\vec{v}(\vec{x},t)$ any of the previously considered bulk
velocity fields,
without  going into  apparent contradictions.
Specifically, the equation (7) with
$\vec{v}(\vec{x},t)\leftrightarrow \vec{b}(\vec{x},t)$   must be
valid.

By resorting to velocity fields $\vec{v}(\vec{x},t)$  which obey
$\triangle \vec{v}(\vec{x},t)=0$, we may pass from (7) to  an equation
of the Euler form, (10), provided (8) holds true  and then
the right-hand-side
of (7) involves a bounded from below effective potential $\Omega $.

An additional requirement is that
$${\vec{F}-{1\over {\rho }}\vec{\nabla }P \doteq  \vec{\nabla }
\Omega \enspace .}\eqno (14)$$

Clearly, in case of a constant pressure we are left with the dynamical
constraint ($\vec{b}\leftrightarrow \vec{v}_E$):
$${\partial _t \vec{b} + (\vec{b}\cdot \vec{\nabla })\vec{b}
= \vec{F}=\vec{\nabla }\Omega } \eqno (15)$$
combining simultaneously the Eulerian fluid and the Markov diffusion
process inputs, \it if and only if \rm
$\vec{F}$ is repulsive, e.g. $- V(\vec{x},t)$ is  bounded from
below.
Quite analogously, by setting $\vec{F}=\vec{0}$, we would get
a constraint
on the admissible pressure term, in view of:
$${\partial _t\vec{b} + (\vec{b}\cdot \vec{\nabla })\vec{b} =
- {1\over \rho }\vec{\nabla }P = \vec{\nabla }\Omega \enspace .}
\eqno (16)$$

Both, in cases  (15), (16) the effective potential $\Omega $ must
respect the functional dependence (on a forward drift and its
 potential)  prescription (8). In addition, the Fokker-Planck
 equation (4) with the forward drift $\vec{v}_E(\vec{x},t) \doteq
 \vec{b}(\vec{x},t)$ must be valid
 for the density $\rho (\vec{x},t)$.

To our knowledge,  in the literature there is known only one specific
class of Markovian diffusion
processes that would render the right-hand-side of Eq. (10)
repulsive but nevertheless account for the troublesome  Newtonian
accelerations, e.g. those of the from $- \vec{\nabla }V$, with $+V$
bounded from below.
Such processes  have forward
drifts that for each suitable,  bounded from below function
$V(\vec{x})$ solve the nonlinear  partial differential equation:
$${\partial _t\vec{b} + (\vec{b}\cdot \vec{\nabla })\vec{b} =
- D\triangle \vec{b} +
\vec{\nabla }(2Q - V)}\eqno (17)$$
with the compensating pressure term:
$${Q \doteq 2D^2 {\triangle \rho ^{1/2}\over \rho ^{1/2}} \doteq
{1\over 2}\vec{u}^2 + D\vec{\nabla }\cdot \vec{u}}\eqno (18)$$
$$\vec{u}(\vec{x},t)= D\vec{\nabla } ln\, \rho (\vec{x},t)$$
Their exhaustive discussion can be found in Refs.
\cite{nel,burg,olk,zambr}, together with indications for their
possible relevance  as a stochastic counterpart of the
Schr\"{o}dinger picture quantum dynamics.
Clearly,  we have:
$${\vec{F}=-\vec{\nabla }V\, , \,
\vec{\nabla }2Q = - {1\over \rho }\vec{\nabla }P}\eqno (19)$$
where:
$${P(\vec{x},t)= - 2 D^2 \rho(\vec{x},t)\, \triangle \,
ln\, \rho (\vec{x},t)}\eqno (20)$$
Effectively, $P$ is here defined up to a
time-dependent constant.
Another admissible form of the pressure term reads (summation
convention is implicit):
$${{1\over \rho }\vec{\nabla }_k[\rho \,  (2D^2 \partial _j
\partial _k)
ln\, \rho ]  =
\vec{\nabla }_j (2Q)}\eqno (21)$$.

If we consider a subclass of processes for which the dissipation term
identically vanishes ( a number of examples can be found in Refs.
\cite{olk}):
$${\triangle \vec{b}(\vec{x},t)=0}\eqno (22)$$
the equation (17) takes a conspicuous Euler form  (10), $\vec{v}_E
\leftrightarrow  \vec{b}$.

Let us notice that  (20), (21) provide for a  generalisation  of
the more
 familiar, thermodynamically motivated and suited for ideal gases and
 fluids,   equation of
 state $P \sim \rho $.  In case of density fields  for which
 $-\triangle ln\, \rho \sim const$, the standard relationship between
 the pressure and the density is reproduced.
 In case of density fields obeying
 $-\triangle ln\, \rho =0$, we are left with at most  purely time
 dependent  or a constant pressure. Pressure profiles may be highly
 complex for arbitrarily chosen initial density and/or the
 flow velocity fields.

To conclude the present discussion let us invoke Refs.
\cite{marra,spohn,olk}.
  The problem of a   diffusion process interpretation
of various partial differential equations   has been extended
beyond the original  parabolic equations setting, to nonlinear
velocity field equations like the Burgers   one, see e.g. \cite{burg}.
On the other hand, the nonlinear Markov processes associated with
the Boltzmann equation,
in the hydrodynamic limit, are known to imply either an ordinary
differential equation
with the velocity field solving the Euler equation, or  a diffusion
process
whose drift is a solution of the incompressible Navier-Stokes
equation
(without the $curl \vec{v}=0$ restriction), \cite{spohn,marra}.
The case of external forcing has never  been satisfactorily solved.

Our reasoning went otherwise.  We asked for the admissible space-time
dependence  of general velocity fields that are to play the r\^{o}le
of
forward drifts  of Markovian diffusion processes. Our finding is that
solutions of the compressible Euler equation  are appropriate for
the description of a non-deterministic  (e.g. random and Markovian)
evolution and belong to a class of Markovian diffusion processes
orginally introduced by E. Nelson in his quest for  a
probabilistic counterpart of the quantum dynamics, \cite{nel,olk}.
Our solution of the problem involves only the gradient velocity
fields.
However, a couple of issues concerning the $curl \, \vec{b} \neq 0 $
velocity
fields and their nonconservative forcing have  been raised
in Refs. \cite{burg}.

\end{document}